# Reward-modulated learning using spiking neural networks for vehicle lateral control

Javier Pérez [1], Manuel A. Vargas [1], Juan A. Cabrera [1], Juan J. Castillo[1], Barys Shyrokau[2].
[1] University of Málaga, Spain; [2]Delft University of Technology, The Netherlands

E-mail: javierperez@uma.es

This paper presents a vehicle lateral controller based on spiking neural networks capable of replicating the behavior of a model-based controller but with the additional ability to perform online adaptation. By making use of neural plasticity and thanks to reward modulation learning, neural connections are modified to adjust the response according to the committed error. Therefore, the error performs a similar role to dopamine in a biological system, modulating the learning process based on spiking time dependency. The connections are initially set to replicate behavior of the model-based controller. Online adaptation allows tuning connection parameters to improve controller performance. A path controller with a preview is used as a baseline controller to evaluate the performance of the proposed approach. Key performance indicators are obtained from simulation with a step response and a handling track with 20 curves of different radii.

Topic / Autonomous Driving Systems

## 1. INTRODUCTION

Spiking neural networks (SNNs), considered the third generation of artificial neural networks, offer the potential to emulate biological learning mechanisms. This characteristic allows SNNs to solve complex control tasks [1] while adding adaptation capabilities.

This study proposes to implement this kind of neural network to perform lateral control of vehicle dynamics, achieving online adaptation.

The use of neural networks as a controller presents an initialization problem of neural connections (synapses) since controller stability depends on its settings. To avoid random initialization of neural connections, it has been proposed to reproduce the behavior of a model-based controller as a baseline. Thus, the proposed controller replicates the baseline controller initially but after a few iterations, its connections are optimized to improve overall performance while preventing stability problems.

Although a predefined structure simplifies learning, if the structure is large [2], it can have the opposite effect, thus increasing the complexity of learning. For example, a human being driving a vehicle could be implemented as the baseline controller, performing behavior cloning [3]. However, since the rules adopted by humans are not directly known, they would have to be extracted using a large and complex network, requiring a large amount of data and computational costs.

Therefore, instead of using a human as a baseline controller, a state-of-the-art controller is used [4]. By using a simpler controller with the ability to adapt [5], it is possible to obtain responses comparable to control algorithms of greater complexity and computation, such as the commonly used model predictive control [6]. Furthermore, adding an adaptive capability to a predictive type of controller [7] makes fulfilling real-time requirements challenging.

Hence, the utilization of spiking neural networks with a prefixed structure [8] and reward modulation-based learning [9] provides online learning [10] and flexibility to the controllers with a simple predefined structure. Thus, it is not necessary to add a predictive control.

A path controller with a preview has been chosen as a baseline controller due to its simple control function (1) and has proven to be robust. With the help of neural adaptation, the response of this controller is enhanced.

The performance of the proposed controller has been improved compared to the baseline controller during the tracking of a route of long duration. Initially, both provide the same performance as the baseline neural network replicates behavior. However, as reward modulation trains the controller, error is gradually reduced.

The main contributions of this work are summarized as follows:

- Neural network model-based controller.
- Spiking neural network pre-defined structure.
- Online learning using reward-modulated learning.
- Performance comparison using a validated multi-body vehicle model.

The paper is organized as follows. In Section 2, the model-based controller implemented as the baseline is introduced. The neural network and learning procedure are presented in Section 3. The result and a discussion of the evaluation of the performance of the proposed controller are included in Section 4. Finally, conclusions are drawn in Section 5.

## 2. MODEL-BASED CONTROLLER

As a baseline controller, a controller of reduced complexity is required although it should be able to have preview capabilities. The chosen controller is a path control with a preview (PCwP). Due to its simplicity, the neural network needs a reduced number of neurons to replicate its behavior, which reduces its computational cost and makes it easier to implement it in an embedded system.

Path control with a preview has a nonlinear control function that determines the steering angle ($\delta$). The function is expressed using a feedback (*fb*) term and a feedforward (*ff*) term (1).

$$\delta(t) = \delta_{fb}(t) + \delta_{ff}(t) \quad (1)$$

A single-track bicycle model is used to determine the control parameters that define both terms. In this case, four variables define state vector **x** (2), which has to be measured or estimated in order to be implemented by the PCwP. Consequently, lateral velocity ($\dot{y}$) and yaw rate ($\dot{\psi}$) as well as the committed lateral ($e_y$) and heading ($e_\psi$) errors have been selected.

$$\boldsymbol{x} = [\dot{y}\ \dot{\psi}\ e_y\ e_\psi]^T \quad (2)$$

Using the state vector, the two PCwP terms are defined according to equation (3) and (4).

$$\delta_{fb}(t) = [0\ 0\ k_{ye}(u)\ k_{\psi e}(u)\ ]\boldsymbol{x} \quad (3)$$
$$\delta_{ff}(t) = (L + k_{ff}(u))/R \quad (4)$$

Where $k_{ye}, k_{\psi e}$ and $k_{ff}$ are the longitudinal speed dependent gains of the controller that needs to be tuned according to the dynamics of the vehicle. The feedforward term is set as a function of the curvature radius (R) of the trajectory, the characteristic velocity gain ($k_{ff}$) and the wheelbase (L).

Encoding the control variables ($e_y, e_\psi, R, u$) and implementing equation (1) in the neural network connections are required to reproduce the PCwP nonlinear function. Gaussian bells are used to encode the input variables whereas to implement the equation, as it is a non-linear function of the input variables, non-linear neural connections are resorted to. The following section is devoted to defining the neural network by making use of axo-axonic connections in addition to the commonly applied axodendritic connections.

## 3. SPIKING NEURAL NETWORK

Leaky Integrate and Fire (LIF) is used as the neural model due to its low computational cost and high firing rates. Equation (5) defines the differential equation that establishes the potential level of the membrane ($u$) as a function of the input current coming from the presynaptic neurons ($c$). Once the membrane potential has reached a certain threshold ($u_{th}$), firing is assumed and its potential is restarted (6).

$$\dot{u} = \tau(c - u) \quad (5)$$
$$if\ u > u_{th}\ then\ u = 0 \quad (6)$$

Where $\tau$ is the time constant of the first-order system.

The learning process is reward-modulated, using the error committed for the modulation of dopamine levels at the neuronal connections. This implies a change in neural connections, typically referred to as neural plasticity. The mechanism selected (7) is Spike-Timing-Dependent Plasticity (STDP).

$$STDP = e^{\frac{-(t_{pre}-t_{post})}{T}} \quad (7)$$

This method makes use of time differences between the presynaptic ($t_{pre}$) and postsynaptic ($t_{post}$) neurons to determine the variation of each connection. Hebbian learning is applied, so neurons with similar firing rates are potentiated as a function of rate constant (T).

As mentioned before, axodendritic ($u^d$) and axo-axonic ($u^a$) connections are utilized to reproduce nonlinear behavior. Both are connected through a nonlinear synapse model (8).

$$c = (u^d)\boldsymbol{W}(u^a) \quad (8)$$

Unlike a linear synapse that has only one input, typically dendritic, the proposed synapse has two inputs, axodendritic and axo-axonic. They multiply the weight matrix of front and back connection weights (**W**) respectively. Thus, a non-linear relationship is established between the input of each presynaptic and postsynaptic neuron. In addition, in order to allow the output to have positive and negative signs, two connection matrixes are required, an excitatory one ($\boldsymbol{W}_e$) associated with the positive output and an inhibitory one ($\boldsymbol{W}_i$), associated with the negative output.

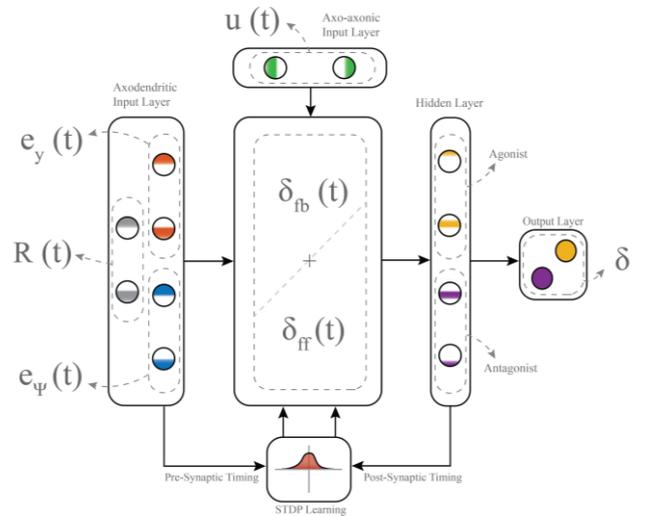

Fig. 1. PCwP Neural Network Structure

Each matrix is associated with a set of neurons in the hidden layer, assigning the excitatory matrix to the agonist neurons and the inhibitory one to the antagonist neurons. Thus, the output representing the steering angle is defined according to equation (9) as a sum of all hidden neuron activity.

$$\delta = (u^d)\, \boldsymbol{W}_e\, (u^a) - (u^d)\, \boldsymbol{W}_i\, (u^a) \qquad (9)$$

The neural network in charge of replicating PCwP behavior is therefore structured as shown in Fig.1. Longitudinal speed is set as the only input of the axo-axonic layer since it is responsible for nolinear behavior of the PCwP while the other variables constitute the axodendritic layer.

A total of 20 neurons are assigned to each variable. Thus, a total of 60 neurons in the axodentric layer and 20 in the axo-axonic layer compose the input layer of the network. The hidden layer accounts for 40 neurons and the output layer for only two neurons adding up to a total of 122 neurons. Therefore, each weight matrix establishes 60x20=1200 neuronal connections. Since an excitatory and inhibitory matrix are required, 2400 weights are initialized to replicate non-linear behavior of the PCwP function. This neural network is integrated into the vehicle control scheme as shown in Fig. 2.

## 4. RESULTS

To evaluate the performance of the proposed algorithm, IPG CarMaker software was used with a validated multi-body vehicle model based on field tests. The performance was evaluated both for a step response and a complex road profile. The corresponding road profile was the handling track of a commercial test field. It has a length of 4.3 km where 20 curves of different radii were used to simulate various driving scenarios.

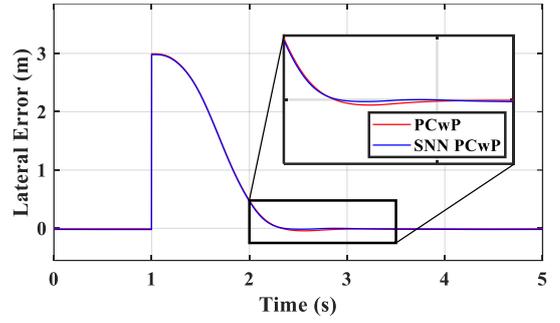

Fig. 3. Lateral error for the step response

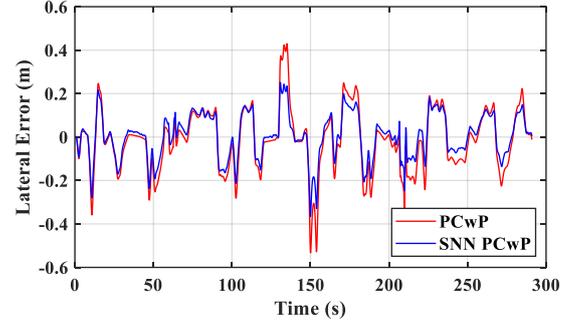

Fig. 4. Lateral error across the handling track

In order to compare the proposed adaptive algorithm (SNN PCwP) with the baseline, key performance indicators were obtained (Fig. 5). For the step response the peak, overshoot, undershoot, rise, and settling time values were chosen. For the handling track, tracking performance was assessed by the root-mean-square error (RMSE) and the maximum (MAX), mean (MEAN) and standard deviation (SD) of the lateral error. Additionally, the lateral jerk (JERK) was used to compare the aggressiveness of both controllers.

In both cases, it can be observed that the result of the proposed adaptive controller outperforms the baseline controller. The step response, although the error difference is very low (Fig. 3), demonstrates the adaptive capability of the neural controller.

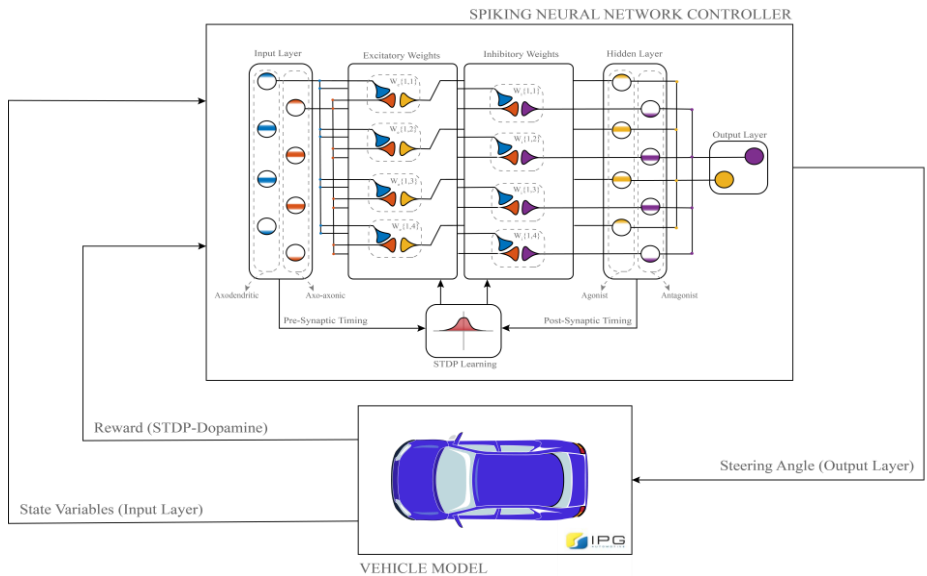

Fig. 2. Vehicle Control Scheme

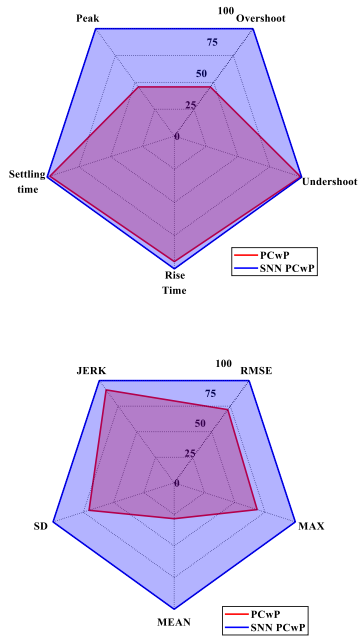

Fig. 5. KPIs for the step response (top) and handling track (bottom)

Even though the response of the baseline controller is optimal, the SNN reduces the lateral error experienced even more, achiving smaller overshoot. For the case of driving on a track, the response is suboptimal as the controller must act correctly for all types of road curvatures. Learning adaptation is shown in Fig.4 where it can be observed that the SNN adapts to the conditions by continuously minimizing the error.

## 5. CONCLUSION

This study proposes to incorporate adaptive capabilities to a state-of-the-art model-based control algorithm. To this end, spiking neural networks with reward learning using spike-timing-dependent plasticity have been developed. Validation of the proposed algorithm has been performed for several scenarios (a step response and a handling track) using several key performance indicators. It has been demonstrated that spiking neural networks can improve controller adaptability and tracking performance without a considerable increase of computational costs.

Future works will be devoted to implementing other model-based controllers that will take into account a larger number of state variables, providing greater flexibility to the neural network. Thus, the error rate will be reduced although the learning process becomes slower as the number of neural connections increases.

Finally, the implementation of the algorithm in an embedded system is required to validate its execution in a real-time system and to be able to control a real vehicle.